\documentclass[aps, prd, twocolumn, showpacs, superscriptaddress, groupedaddress]{revtex4} 
\usepackage{graphicx}    
\usepackage{amssymb}
\usepackage[normalem]{ulem}
\usepackage{subfigure}
\usepackage{float}
\usepackage[normalem]{ulem}
\usepackage[pagebackref=false, colorlinks=true]{hyperref}
\hypersetup{
linkcolor=blue,     % color of internal links
citecolor=blue,     % color of links to bibliography
urlcolor=blue}   
\usepackage{soul}
\usepackage{amsmath}
\begin{document}
\title{Shadow Formation Conditions Beyond the Kerr Black Hole Paradigm}
\author{Parth Bambhaniya}
\email{grcollapse@gmail.com}
\affiliation{Instituto de Astronomia, Geofísica e Ciências Atmosféricas, Universidade de São Paulo, IAG, Rua do Matão 1225, CEP: 05508-090 São Paulo - SP - Brazil.}

\author{Saurabh}
\email{saurabh@mpifr-bonn.mpg.de}
\affiliation{Max-Planck-Institut f\"ur Radioastronomie, Auf dem H\"ugel 69, D-53121 Bonn, Germany}

\author{Elisabete M. de Gouveia Dal Pino}
\email{dalpino@iag.usp.br}
\affiliation{Instituto de Astronomia, Geofísica e Ciências Atmosféricas, Universidade de São Paulo, IAG, Rua do Matão 1225, CEP: 05508-090 São Paulo - SP - Brazil.}

% \date{\today}

\begin{abstract}
A compact object illuminated by background radiation produces a dark silhouette. The edge of the silhouette, or shadow, or the apparent boundary, or the critical curve is commonly determined by the presence of the photon sphere (or photon shell -- in the case of rotating spacetime), corresponding to the maximum of the effective potential for null geodesics. While this statement stands true for Kerr black holes, here we remark that the apparent boundary (as defined by Bardeen \cite{Bardeen}), forms under a more general condition. We demonstrate that a shadow forms if the effective potential of null geodesics has a positive finite upper bound and includes a region where photons are trapped or scattered. Our framework extends beyond conventional solutions, including (but not limited to) naked singularities. Furthermore, we clarify the difference between the apparent boundary of a dark shadow, and the bright ring on the distant observer's screen. These results provide a unified theoretical basis for interpreting observations from the Event Horizon Telescope (EHT) and guiding future efforts toward extreme resolution observations of compact objects.\\\\
$\boldsymbol{Key words}$ : Shadow, Photon sphere, Naked singularities, Black Holes,  EHT.
\end{abstract}
\maketitle

\section{Introduction}

The Event Horizon Telescope (EHT) collaboration has made remarkable strides in exploring the evidence for the event horizon of a black hole, significantly advancing our understanding and opening new avenues for other possibilities. In contrast, as outlined in recent EHT analyses (see, for example~\cite{EventHorizonTelescope:2022xqj}), the results and analyses based on
metric tests, suggested that observational evidence does not definitively rule out the possibility that Sgr~A* could be a naked singularity, a compact object without an event horizon. 
Thus, the analyses favors an open and careful interpretation of their findings. This ongoing uncertainty not only spurs scientific debate but also attracts the attention of both the scientific community and the general public, highlighting the profound implications of these findings.  

Ultra-compact objects such as singular black holes, regular black holes, charged black holes, naked singularities, wormholes, boson stars and gravastars are key to understanding the shadows and accretion disk properties within the general relativity and other gravity theories \cite{Vagnozzi:2022moj}. Theoretical studies of gravitational collapse in general relativity suggest that both black holes and naked singularities can emerge, depending on initial conditions and physical parameters \cite{joshi,goswami,mosani1,mosani2,mosani3,mosani4,Deshingkar:1998ge,Jhingan:2014gpa,Joshi:2011zm}. One of the most critical aspects of studying such objects is understanding their observational signatures, particularly in light of recent and upcoming Event Horizon Telescope (EHT) observations \cite{EventHorizonTelescope:2019dse,EventHorizonTelescope:2019uob,EventHorizonTelescope:2022xnr,EventHorizonTelescope:2024hpu,EventHorizonTelescope:2024rju,Ayzenberg:2023hfw}. The detection of the brightness depression in the images of M87* and Sgr~A* by the EHT has provided a direct way to probe the nature of compact objects and test general relativity in the strong-field regime. Although, different ultra-compact objects can cast distinct shadow structures, depending on their spacetime geometry and photon trajectories \citep[see for e.g.,][]{EventHorizonTelescope:2020qrl}.
% \cite{shaikh1,Solanki:2021mkt,Saurabh:2022jjv,Bambhaniya:2024hzb,gyulchev,ohgami_2015,Sakai,Bambhaniya:2021ugr,Vertogradov:2024fva,Bambhaniya:2024lsc,Bambhaniya:2021ybs,ABJoshi}. 

These variations in shadow morphology necessitate a framework that can systematically characterize the shadow formation (under the definition mentioned above) across diverse compact object cases.

Deriving generalized shadow formation conditions is essential for distinguishing between black holes, naked singularities, and other exotic compact objects in future high-resolution observations. The shadow characteristics, such as size, shape, and brightness distribution, could provide direct evidence for the existence of horizonless objects, thereby challenging the cosmic censorship conjecture and offering deeper insights into quantum gravity and modified theories of gravity. Future observations with the EHT, combined with theoretical advancements in shadow analysis, could play a crucial role in unveiling the true nature of these ultra-compact objects.

Recently Broderick et al. \cite{Broderick:2024vjp}, demonstrated that a broad class of naked singularities exhibit inner turning points for timelike geodesics in the parameter space when shadow occurs, resulting in the presence of an accretion-powered photosphere inside the shadow region of the naked singularity. This implies that the presence of any accretion shock must appear inside the photon sphere if it is a naked singularity. However, the shadows of Sgr~A* and M87* in the EHT observations, indicate that the accretion flow remains well ordered down to the photon sphere. Therefore, a broad class of naked singularities, except JMN-1 and JNW naked singularities (within some parameter space), can be generically excluded \cite{Broderick:2024vjp}. Since these naked singularities do not have generic turning points for timelike geodesics prior to reaching the singularity, it will be difficult to observe accretion-driven shock or photosphere within the shadows of JMN-1 and JNW naked singularities using this approach. In other words, unlike other naked singularities, JMN-1 and JNW spacetimes lack generic turning points for timelike geodesics before reaching the singularity, meaning accretion-driven shocks or photospheres do not develop outside the singularity. As such, they are not ruled out by current EHT shadow constraints.

Prior work demonstrates that the shadow forms due to the presence of a photon sphere in the spacetime geometry of the ultracompact object (See, for example, ~\cite{Saurabh:2023otl,shaikh1}). However, it cannot define any fundamental relation between the photon sphere, an event horizon, and the singularity to form a shadow. Notably, the presence of a photon sphere will not confirm the simultaneous existence of an event horizon and a singularity. Here, classical singularity is referred to as the boundary of a spacetime manifold where non-spacelike geodesics (timelike and null) are incomplete and volume element of Jacobi field vanishes infinitesimally. An event horizon is a boundary of a causal past of future null infinity, and a photon shell (in the case of axisymmetric rotating spacetimes) is the region of unstable photon orbits of null geodesics. The key interesting point is that a shadow can form whether the singularity is absent, hidden behind the event horizon, or naked, for example, regular BHs, singular BHs and naked singularities, wormholes \cite{Vagnozzi:2022moj}. 

A recent study showed that a shadow can be cast without a photon sphere, in this case the singularity itself casts a shadow, but with smaller and larger diameters \cite{Joshi:2020tlq,Dey:2020bgo}. This is justified by the fact that shadow formation fundamentally depends on the causal structure of null geodesics and the presence of regions where photons are either trapped or scattered. The effective potential provides a covariant way to characterize these regions across a wide class of spacetimes. While in Kerr black hole, the photon sphere is associated with a peak in the effective potential. We show that even in geometries without a photon sphere, a finite upper bound of the effective potential is sufficient for a shadow to form. Our formalism accommodates both cases, offering a unified description of shadow formation beyond specific solutions. While some of previous studies \cite{Joshi:2020tlq,Vagnozzi:2022moj}, including some by the present authors, have investigated shadow formation for compact objects such as naked singularities and regular black holes. These works have largely focused on specific analytic spacetimes, most commonly spherically symmetric and static geometries. For instance, shadow formation in naked singularities has typically been studied in the context of null or timelike singularities in non-rotating metrics.

In contrast, the novelty of the present work lies in formulating a generalized, model-independent condition for shadow formation that applies to \emph{axisymmetric}, \emph{stationary}, and \emph{rotating} spacetimes. Our approach is based entirely on the behavior of the effective potential for null geodesics, and we show that shadow formation is governed by the existence of a finite potential barrier capable of trapping or scattering photons regardless of the presence of a conventional photon sphere or event horizon. This framework thus unifies and extends earlier results \cite{Joshi:2020tlq,Vagnozzi:2022moj} by providing necessary and sufficient conditions for the existence of a shadow in a wide class of spacetimes, many of which are relevant for ongoing and future observations by instruments such as the Event Horizon Telescope (EHT). To the best of our knowledge, such a general condition for silhouette formation in rotating, axisymmetric spacetimes has not been previously presented in the literature.

Given these diverse cases where the shadow is formed in the context of Bardeen like definition for Kerr spacetime \cite{Bardeen}, we lay out a generalized condition to form a shadow for axisymmetric stationary and rotating spacetime using the effective potential of null geodesics in Section~\ref{secII}. This also incorporates the instability of a photon sphere with varying spin parameter to build shadow formation conditions. In Section~\ref{secIII}, we discuss our conclusions. Throughout this paper, we use the gravitational constant $G$ and the speed of light $c$ as a unit value. 

\section{Silhouette formation conditions}
\label{secII}
Let $(M, g_{\mu\nu})$ represent a general axisymmetric, stationary, and rotating spacetime with a metric signature (-,+,+,+). The general rotating spacetime metric is described by,
\begin{equation} \label{eqn:1}
    ds^2=g_{tt}dt^2+g_{rr}dr^2+g_{\theta\theta}d\theta^2+g_{\phi\phi}d\phi^2+2g_{t\phi}dtd\phi,
\end{equation}
here, $g_{tt}, g_{rr},g_{\theta\theta}$, $g_{\phi\phi}$ and $g_{t\phi}$ are metric tensor components. The metric in Eq. (1) represents a general stationary, axisymmetric, and rotating spacetime. We assume that the spacetime admits two Killing vectors corresponding to time translation and axial rotation symmetries, allowing conserved energy and angular momentum for geodesics. This formulation includes both vacuum and non-vacuum spacetimes and does not assume asymptotic flatness or regularity at the origin. However, our framework is limited to spacetimes where the geodesic equations remain well-defined and separable (at least in the equatorial plane), and excludes non-axisymmetric or time-dependent metrics, such as perturbed rotating systems or binary black hole mergers. Additionally, we assume that the metric functions are smooth and differentiable in the domain of interest and that the effective potential for null geodesics remains finite. For axisymmetric stationary spacetimes, we have the following conserved quantities due to temporal and axial symmetries:
\begin{equation}
    E=-P_t=-g_{tt}\frac{dt}{d\lambda}-g_{t\phi}\frac{d\phi}{d\lambda},
\end{equation}
\begin{equation}
    L_z=-P_{\phi}=g_{\phi\phi}\frac{d\phi}{d\lambda}+g_{t\phi}\frac{dt}{d\lambda}.
\end{equation}
were $P^\mu$ is the four-momentum of the null geodesic, $E$ is the photon's energy and $L_z$ is the photon's angular momentum measured at infinity. The radial motion of photons is governed by the null condition (that the tangent vector be null),
\begin{equation}
    g_{\mu\nu}P^{\mu}P^{\nu}=0.
\end{equation}
Substituting the expressions for conserved quantities into the null geodesic condition, the radial motion equation becomes:
\begin{multline}
    g_{rr}\left(\frac{dr}{d\lambda}\right)^2=-g_{tt}\left(\frac{dt}{d\lambda}\right)^2-2g_{t\phi}\left(\frac{dt}{d\lambda}\right)\left(\frac{d\phi}{d\lambda}\right)\\- g_{\phi\phi}\left(\frac{d\phi}{d\lambda}\right)^2-g_{\theta\theta}\left(\frac{d\theta}{d\lambda}\right)^2,
\end{multline}
The radial motion of null geodesics can be expressed using an effective potential formulation as:
\begin{equation}
\frac{1}{2}\left(\frac{dr}{d\lambda}\right)^2 + \mathcal{V}_{\text{eff}}(r) = 0,
\end{equation}
where $\mathcal{V}_{\text{eff}}(r)$ denotes the effective potential governing photon orbits. It encodes the combined influence of the spacetime geometry and conserved quantities on the radial motion of photons. In spherically symmetric spacetimes, the photon sphere is a spherical surface composed of unstable circular photon orbits at a fixed radius $r = r_{\text{ph}}$. However, in axisymmetric rotating spacetimes (such as Kerr), this generalizes to a \emph{photon shell}, a region spanning a range of radii where unstable bound photon orbits exist. This shell contains distinct prograde and retrograde circular photon orbits due to the effects of frame dragging. The presence of a photon shell results in an asymmetric shadow boundary, especially for high spin parameters. Here, $\mathcal{V}_{\rm eff}(r,\theta)$ can be reduced as a function of $r$ only on equatorial plane because photon sphere corresponds to circular photon orbits with constant $\theta$. However, here the effective potential is defined in a generalized manner,
\begin{multline}
   \mathcal{V}_{\rm eff}(r)=\frac{1}{2g_{rr}}( g_{tt}\left(\frac{dt}{d\lambda}\right)^2+2g_{t\phi}\left(\frac{dt}{d\lambda}\right)\left(\frac{d\phi}{d\lambda}\right)\\+ g_{\phi\phi}\left(\frac{d\phi}{d\lambda}\right)^2),
\end{multline}
\begin{figure*}
    \centering
    \includegraphics[width=\textwidth]{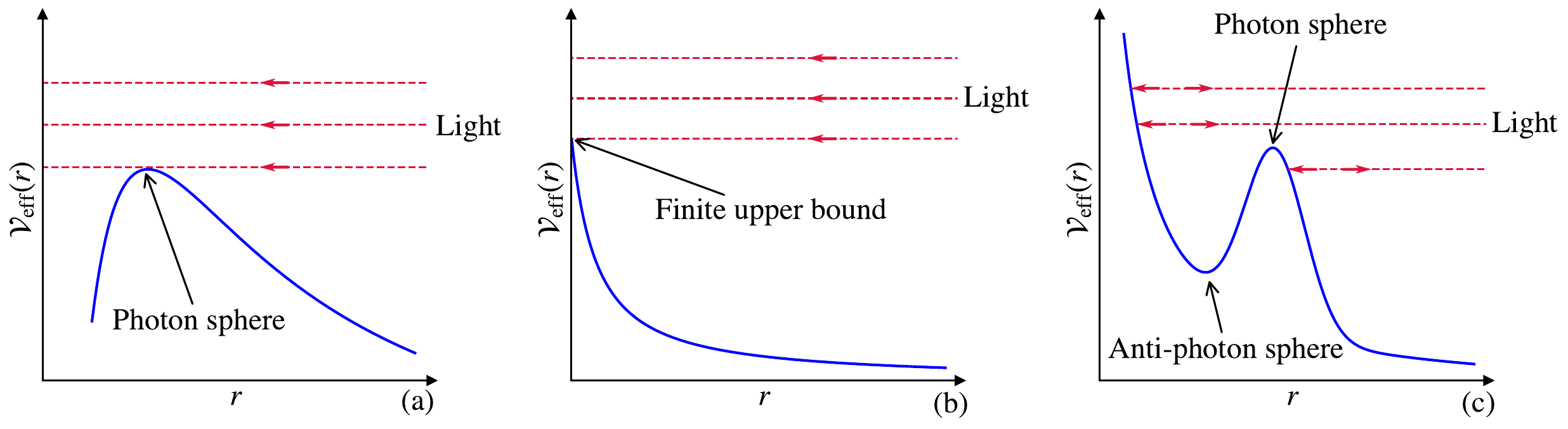}
    \caption{Representation of various (but not limited to) effective potentials. Here (a) A finite upper bound with a photon sphere, typical of Schwarzschild or Kerr black holes, JMN-1 and JNW naked singularities with photon spheres. 
(b) A finite upper bound without a photon sphere, possible in certain naked singularity or wormhole geometries. 
(c) A photon sphere exists but the potential diverges, representing geometries with pathological energy behavior (e.g., anti-photon sphere). In rotating spacetimes, the photon sphere generalizes to a photon shell, a region of unstable photon orbits distributed over a finite radial range. This leads to the formation of asymmetric shadow boundaries as observed in Kerr spacetimes.}
    \label{fig:veff}
\end{figure*}
Physically, photon trapping regions can be thought of as gravitational potential wells where photons are temporarily confined due to the curvature of spacetime. In the case of a photon sphere (as in Figure~\ref{fig:veff}(a)), photons can orbit in unstable circular trajectories, and small perturbations cause them to either escape to infinity or plunge inward. Even when such a sphere is absent, as in Figure~\ref{fig:veff}(b), the presence of a finite potential barrier means that photons can still be deflected or delayed leading to a decrease in observed intensity. This mechanism underlies the formation of shadows even without sharp ring features. Hence, the behavior of the effective potential governs not just photon orbits but also the visibility structure of the shadow region.

The generalized condition for shadow formation should follow the below
\begin{itemize}

    \item  Non-divergence of effective potential: The effective potential must remain finite and continuous throughout the region and should not diverge positively near the horizon or singularity. Even if it has a photon sphere or photon shell as a special case this should hold true. The conditions for determining unstable circular orbits (photon sphere/shell) is given by
    \begin{equation} \label{eqn:10}
         \mathcal{V}'_{\rm eff}(r)=0; \mathcal{V}_{\rm eff}''(r)<0;
    \end{equation}
    \begin{equation}
        \mathcal{V}_{\rm eff}(r)<\infty;     \forall (r<r_{\rm ph}).
    \end{equation}
    
    Here, $r_{\rm ph}$ is the radius of the photon sphere and a prime denotes the derivative with respect to $r$. This condition excludes singularities or regions of unbounded energy that could disrupt shadow formation. 
\end{itemize}
    Physically, a divergent effective potential signals pathological behavior in the spacetime, such as infinite energy densities or curvature singularities, where photon trajectories cannot be meaningfully resolved. In such scenarios, photon motion ceases to be predictable, and a clear separation between escaping and captured photon trajectories becomes impossible. Consequently, there is no well defined critical impact parameter or shadow boundary. Therefore, requiring the effective potential to be finite ensures that null geodesics remain regular and evolve predictably, making the formation of a photon trapping region and hence a shadow formation possible, even in the absence of a conventional photon sphere.
    
    Figure~\ref{fig:veff} illustrates representative cases of the effective potential $\mathcal{V}_{\rm eff}(r)$ for a given value of $L_z$. Figures~\ref{fig:veff}(a) and~\ref{fig:veff}(b) depict scenarios where shadow formation occurs with and without an associated photon sphere, respectively. Specifically, the presence of a bright and sharply delineated photon ring in the shadow image is associated with an effective potential resembling Figure~\ref{fig:veff}(a), while a shadow characterized by a brightness depression without a pronounced ring structure corresponds to a potential of the type shown in Figure~\ref{fig:veff}(b). This encapsulates the cases where the shadow images does not see the presence of bright ring as a consequence of the increased optical path of the photon through the optically thin plasma. 
    Then, provided spacetime metric by equation~\ref{eqn:1}, the critical boundary or the shadow radius (in the equatorial plane) is then given by solving equation~\ref{eqn:10}. The shadow boundary corresponds to the critical photon orbits, defined by the critical impact parameter,
\begin{equation}
b_{\text{crit}} \equiv \frac{L_z}{E},
\end{equation}
where $L_z$ and $E$ are the conserved angular momentum and energy of the photon, respectively. This parameter determines the apparent size and shape of the shadow observed at infinity. For a given metric, $b_{\text{crit}}$ satisfies:
\begin{equation}
b_{\text{crit}} = \frac{-g_{t\phi} \pm \sqrt{(g_{t\phi})^2 - g_{tt}g_{\phi\phi}}}{g_{tt}}.\label{12}
\end{equation} where, $b_{\rm crit}$ is critical impact parameter and defines the shadow boundary with $r=r_{\rm ph}$. Physically, the critical impact parameter $b_{\text{crit}} \equiv L_z/E$ defines the threshold between photon trajectories that are captured and those that escape to infinity. In rotating spacetimes such as Kerr, this boundary is direction-dependent due to frame dragging, leading to asymmetric shadows. Equation (\ref{12}) provides the condition for the circular photon orbits (shadow boundary) and encodes how the spin of the compact object distorts the shadow through metric components $g_{tt}, g_{t\phi},$ and $g_{\phi\phi}$. 

It is important to note that Eq. (11) assumes that the null geodesic equations are separable, which is the case for several well known axisymmetric, stationary spacetimes including Kerr, Kerr-Newman, and certain regular or naked singularity geometries. In our analysis, we adopt separability at least in the equatorial plane, which is a valid and widely used approximation in astrophysical scenarios. This ensures that the conserved quantities such as $L_z$ and $E$ can be defined in a consistent manner, allowing the critical impact parameter to be computed. The spin parameter $a$ introduces asymmetry in the photon sphere radii and shadow boundary, modifying the critical photon sphere locations and impact parameters $b_{\rm crit}$. In a rotating spacetime (e.g., Kerr-like geometry), the spin parameter modifies the effective potential and the photon sphere structure. The ``photon sphere" is no longer a single valued circular photon orbit situated at $r_{\rm ph}$ but rather a ``shell" with some thickness ranging in $r_{-} \leq r_{\rm ph} \leq r_{+}$. This photon shell, in the case of Kerr BH in equatorial plane, contains two circular photon orbits: one which is co-rotating (prograde) with respect to the BH spin, and the other one is counter-rotating (retrograde). The co-rotating and counter rotating photon orbits are distinct due to frame dragging. As the spin parameter increases $(a\rightarrow M)$, the co-rotating photon orbit becomes smaller and approaches the event horizon, whereas the counter-rotating photon sphere remains farther out. In the observations, the direction of the line of sight fundamentally influences the apparent shape, size, and asymmetry of the shadow.

\section{Discussions and Conclusions}
\label{secIII}
In this work, we have established a comprehensive and generalized conditions for shadow formation in general axisymmetric, stationary, and rotating spacetimes, providing both necessary and sufficient conditions for the appearance of shadows. Our analysis reveals that shadows form if the effective potential of null geodesics possesses a positive finite upper bound and includes a region where photons are either trapped or scattered. By incorporating the effects of spin, we demonstrate how it influences the instability of photon orbits and induces asymmetry in the shadow boundary. Our framework extends beyond traditional black holes, suggesting that under suitable conditions, shadows may form in spacetimes of naked singularities or regular compact objects even in cases lacking conventional photon spheres. However, we emphasize that the existence and observational features of such shadows depend on the behavior of null geodesics and the finiteness of the effective potential. We also provide a qualitative comparison of shadow features for different compact objects. Schwarzschild black holes produce circular, symmetric shadows due to the presence of a single photon sphere, while Kerr black holes show noticeable asymmetry and D-shaped shadows as a result of frame-dragging effects. Naked singularities such as JMN-1 and JNW spacetimes may exhibit either similar and slightly smaller shadow diameters respectively and can produce shadows with or without bright photon rings depending on the presence or absence of a photon sphere \cite{Vagnozzi:2022moj,Saurabh:2022jjv,Saurabh:2023otl}. In cases lacking a photon sphere, the shadow may still form due to photon scattering but often lacks a sharp brightness depression. Regular black holes, wormholes, null and timelike naked singularities may present more diffused or nonstandard shadow edges \cite{Vagnozzi:2022moj,Joshi:2020tlq,Dey:2020bgo}. These qualitative distinctions provide an observational basis for differentiating between black holes and other possible compact objects using high resolution imaging. These results should be interpreted within the assumptions of our framework and may not generalize to all non-Kerr scenarios.

This pedagogical study offer a unifying theoretical foundation that advances the interpretation of observational data from the Event Horizon Telescope (EHT) and guides next-generation imaging missions. Our results will aid in identifying a broader range of compact objects and exotic gravitational geometries, enhancing our understanding of strong-field gravitational physics. The key interesting point is that a shadow can form whether the singularity is absent, hidden behind the event horizon, or naked, for example, in regular black holes, singular black holes, naked singularities, and wormholes. A recent study showed that a shadow can be cast without a photon sphere \cite{Joshi:2020tlq,Dey:2020bgo}, in which case the singularity itself contributes to the shadow formation, leading to variations in shadow size, appearing either smaller or larger depending on the spacetime properties. Therefore, based on the nature of effective potentials of different spacetime geometries as given in Figure \ref{fig:veff}, the various theoretical models of compact objects can be restricted or excluded for shadow imaging.

These findings also have significant implications for future observational studies, particularly with advancements in high-resolution imaging techniques such as the Event Horizon Telescope (EHT). The ability to differentiate between black holes, naked singularities, and other ultra-compact objects based on their shadow properties could provide a new way to test general relativity in the strong-field regime. Notably, the possibility of detecting shadows without photon spheres challenges conventional assumptions and opens new avenues for exploring alternative theories of gravity and quantum gravitational effects. 

Beyond classical spacetime geometries, quantum-gravity inspired scenarios such as dark photon instabilities influenced by the Barbero–Immirzi parameter, or axion–torsion couplings may further modify shadow structures. As shown in \cite{Gao2025}, such effects could lead to deviations in photon trapping behavior or critical curve morphology. Incorporating these quantum corrections into shadow formation frameworks will be essential for testing extensions of general relativity in future ultra-high resolution observations. Future work should focus on refining theoretical models to better predict shadow features for different compact objects and comparing these predictions with observational data.
\\

\acknowledgments{}
We thank Pankaj S. Joshi and Maciek Wielgus for helpful discussions.
P. Bambhaniya and E. M. de Gouveia Dal Pino acknowledge support from the São Paulo State Funding Agency FAPESP (grant 2024/09383-4). E. M. de Gouveia Dal Pino also acknowledges the support from FAPESP (grant 2021w/02120-0) and CNPq (grant 308643/2017-8). Saurabh received financial support for this research from the International Max Planck Research School (IMPRS) for Astronomy and Astrophysics at the Universities of Bonn and Cologne.

\end{document}